\documentclass{aa}
\usepackage{graphics,epsfig,astron,amsmath,amssymb,amstext}

\def\d {\mathrm{d}}
\def\Ep {\hbox{$E^{\prime}$}}
\def\Epp {\hbox{$E^{\prime\prime}$}}

\begin{document}

\thesaurus{12(02.14.1; 09.03.2)}

\title{Time dependent models for the interaction of energetic 
particles in the ISM}

\author{Etienne Parizot}

\institute{Dublin Institute for Advanced Studies, 5 Merrion Square,
Dublin 2, Ireland\\ e-mail: parizot@cp.dias.ie}

\date{(accepted for publication in A\&A)}

\titlerunning{Time dependent models for EP interactions in the ISM}
\maketitle

\begin{abstract}
We present a general model allowing one to calculate the distribution 
function of energetic particles (EPs) in the interstellar medium 
(ISM), and hence any relevant nuclear reaction rate, for any given 
time-dependent injection function, as well as in situations when the 
propagation coefficients themselves are non stationary.  We review and 
provide physical interpretation of general formal solutions of a 
propagation equation taking into account energy losses, nuclear 
destruction and escape.  Both one-zone and extended models are 
investigated.  Our main goal is to provide standard and general tools 
for subsequent use in nuclear astrophysics, notably to calculate the 
gamma-ray emission and spallation rates associated with energetic 
events and specific acceleration mechanisms.  Considering the 
stationary limit of the model, we show that only time-dependent 
calculations can probe the density dependence of the physical 
processes under study, while steady-state models can \emph{in 
principle} only give information about density-independent processes.

\end{abstract}

\section{Introduction}

The energetic particles (EPs) present in the Galaxy can be probed 
through their electromagnetic and nuclear interactions within the 
interstellar medium (ISM).  These interactions can lead to different 
types of radiation, lines (e.g.  from nuclear de-excitation or pair 
annihilation) or continuum (e.g.  synchrotron, bremsstrahlung, inverse 
Compton or $\pi^{0}$ disintegration), as well as to the synthesis of 
secondary nuclei such as Li, Be and B (LiBeB), whose astrophysical 
significance is well established (e.g.  Reeves et al.  1970; Meneguzzi 
et~al.  1971; Fields et al.  1994; Reeves 1994; Ramaty et al.  1996, 
Vangioni-Flam et al.  1996).

Basically, the calculation of spallation and gamma-ray production 
rates at any place in the Galaxy requires the knowledge of i) the 
cross sections for relevant physical processes and ii) the 
distribution function of all the interacting EP species, i.e.  their 
energy spectrum and number density.  In this paper, we indicate how to 
calculate these distribution functions in some general situations 
where an exact formal solution exists.  The different reaction rates 
can then be obtained by merely integrating the corresponding cross 
sections over the EP spectral densities, given the local ISM (target) 
density and chemical composition.  The knowledge of the local magnetic 
field and radiation background further allows one to compute the 
synchrotron and inverse Compton continuum emission in much the same 
way.

In most previous nuclear astrophysics calculations, a steady-state 
assumption has been used, in which the steady injection of EPs in the 
region under study (generally as the result of some particle 
acceleration process) is counterbalanced by a drift in energy space 
due mainly to Coulombian energy losses.  The resulting EP distribution 
functions are thus independent of time, as are consequently all the 
reaction rates.  In this paper, we investigate the case when i) the 
injection of EPs and ii) the conditions of their propagation in the 
surrounding medium are functions of time.  The former type of time 
dependence arises naturally when the acceleration conditions in a 
specific region are changing, for instance when a supernova explodes, 
allowing for acceleration locally in space and time.  As for the 
latter, it can happen when the density and/or composition of the 
ambient medium varies, as for example in an expanding medium such as 
the interior of a supernova remnant (Parizot \& Drury 1999a,b).

In general, a steady-state model is sufficient provided that the time 
scale of the changes in the injection and/or propagation conditions is 
larger than the time needed for the EP distribution function to reach 
its equilibrium, that is, basically, the time scale of energy losses 
or escape out of the region considered.  In this case, the history of 
EP-ISM interactions is fairly well reproduced by juxtaposing different 
phases in which the steady-state approximation holds.  Such a strategy 
has been used extensively for instance to describe the Galactic 
evolution of light element abundances, as induced by GCR interactions 
with the ISM (most recently, Fields \& Olive 1999; Vangioni-Flam et 
al.  1999).  Indeed, the spallation rates at time $t$ depend on the 
current ambient metallicity and the current flux of GCR, which are 
both functions of time, but with time scales presumably larger than 
the GCR confinement time.  There are some situations, however, where 
the steady-state approximation does not hold (Parizot et~al.  
1997a,b,c; Parizot \& Drury 1999a,b), and a time dependent model is 
needed.

It has to be noted that formal solutions of the propagation equation 
have been known for several decades (e.g.  Syrovatskii, 1959; Jones, 
1970; Reames, 1974; Ramaty, 1974) and used extensively in the 
cosmic-ray transport theory, with additional features and 
complications, such as convection, re-acceleration, spatially 
dependent diffusion, etc.  (Lerche and Schlickeiser 1981, 1988; 
Schlickeiser 1986; Wang and Schlickeiser, 1987).  However, the case 
when the energy losses and/or the diffusion coefficient are 
explicitely depending on time, and not only on space, has not yet been 
considered, to the best of our knowledge, while it is relevant to many 
applications in nuclear astrophysics (e.g.  Parizot \& Drury 1999a,b).  
Moreover, most of the situations of interest for spallative 
nucleosynthesis and gamma-ray line astronomy do not require the 
refinements mentioned, so we shall try to summarize and present in a 
clear and unified way the only results which one needs to develop a 
general time-dependent code for nuclear astrophysics calculations.  In 
particular, we shall emphasize the simple and intuitive physical 
meaning of the solutions, describing with special care the more 
original features, especially useful for numerical implementation, 
such as the temporal connection of solutions obtained in different 
phases of the EP propagation.

\section{Basic physical ingredients}
\label{PhysicalIngredients}

The general structure of a spallation or gamma-ray production 
calculation can be divided into three stages : acceleration 
(production of the EPs), propagation (transport in the surrounding 
medium) and interaction (with the surrounding matter or radiation 
field).  These three logical stages are not necessarily separated in 
time, as (re-)acceleration may occur while the EPs are propagating, 
and interactions with the ambient medium just cannot be avoided at any 
stage of the process.  In many situations, however, the acceleration 
time scale of the particles is much shorter than their interaction and 
energy loss time scales, so that acceleration can actually be treated 
simply as \emph{injection} (of EPs).  This means that the acceleration 
itself is not affected by the conditions of propagation, and therefore 
the two stages are disconnected.  In this case, we are solely 
concerned with the future of the EPs once they have been accelerated 
(or advected from a distant acceleration site) and injected into the 
region of space under study.

Accordingly, the first physical ingredient that we need to know is the 
\emph{injection function}, $q_{i}(E,\vec{r},t)$, defined as the number 
of particles of species $i$ introduced (injected) at energy $E$, 
location $\vec{r}$ and time $t$, per unit energy, volume and time 
(in $(\mathrm{MeV/n})^{-1}\mathrm{cm}^{-3}\mathrm{s}^{-1}$).  This 
injection function can be either postulated, in order to reproduce 
specific astronomical observations, or calculated as the output of 
some known acceleration mechanism.

The treatment of the two subsequent stages, propagation and 
interaction, is also made easier by artificially separating them in 
time, which can be legitimated in most cases as discussed below.  In 
any case, the influence of the propagation of the EPs on their 
distribution function can be derived from the knowledge of : i) 
spatial diffusion properties, ii) drift properties in the energy 
space, iii) the local production rate (as secondary nuclei) and iv) 
the `catastrophic' loss rate.  By catastrophic loss we mean the 
outright disappearance of the particle, either by escape out of the 
region under study, destruction in a nuclear reaction (in-flight 
spallation), or radio-active decay in the case of unstable particles.  
This is mathematically described by the total loss time, 
$\tau_{i}^{\mathrm{tot}}(E,\vec{r},t)$, which depends on the nuclear 
species considered, $i$, and is \emph{a priori} a function of energy, 
location and time.  It includes the escape, nuclear destruction and 
decay times according to~:
\begin{equation}
	\frac{1}{\tau_{i}^{\mathrm{tot}}(E,\vec{r},t)} = 
	\frac{1}{\tau_{i}^{\mathrm{esc}}(E,t)} +
	\frac{1}{\tau_{i}^{\mathrm{D}}(E,\vec{r},t)} +
	\frac{1}{\gamma(E)\tau_{i}^{\mathrm{dec}}},
	\label{totalLossTime}
\end{equation}
where $\gamma(E)$ is the Lorentz factor of the particle.

Strictly speaking, the escape time, $\tau_{i}^{\mathrm{esc}}$, is 
relevant only to one-zone models, when the $\vec{r}$ variable is 
meaningless.  Indeed, in spatially extended models, the escape of 
particles out of the region under study is taken automatically into 
account through diffusion, and should not be included in 
$\tau_{i}^{\mathrm{tot}}$.

The nuclear destruction time, $\tau_{i}^{\mathrm{D}}$, is obtained 
from the total inelastic cross sections $\sigma_{i,j}$ for a 
projectile $i$ in a target of species $j$, as~:
\begin{equation}
	\frac{1}{\tau_{i}^{\mathrm{D}}(E,\vec{r},t)} =
	[\sum_{j}\sigma_{i,j}(E)n_{j}(\vec{r},t)]v(E),
	\label{destructionTime}
\end{equation}
where $v(E)$ is the velocity of the particle and $n_{j}(\vec{r},t)$ is 
the local number density of the target nuclei of species $j$, at time 
$t$.

Similarly, the production rate of the EP species $k$ as secondary 
nuclei, through all the nuclear reactions $i + j \rightarrow k$ with 
cross section $\sigma_{i,j;k}(E)$, is given by~:
\begin{equation}
	q_{k}^{\prime}(E,\vec{r},t) = 
	\sum_{i,j}\int_{0}^{\infty}\hspace{-8pt}\d\Ep 
	n_{i}(\Ep,\vec{r},t) n_{j}(\vec{r},t) 
	\sigma_{i,j;k}(\Ep) v_{i}(\Ep),
	\label{SecondaryProductionRate}
\end{equation}
where $n_{i}(E,\vec{r},t)$ is the distribution function of energetic 
nuclei of species $i$, i.e.  their number per unit volume and energy 
(in (MeV/n)$^{-1}$cm$^{-3}$).  Although 
$q_{i}^{\prime}(E,\vec{r},t)$ may be thought of as part of the 
injection function $q_{i}(E,\vec{r},t)$, it should be noted that it is 
actually the part of the injection rate which depends on the EP 
content of the ISM. This makes $q_{i}^{\prime}$ mathematically more 
difficult to implement than the \emph{primary} injection function, 
$q_{i}$, which is the same whatever the local and current distribution 
functions $n_{i}$ may be.

Apart from being created (synthesized) and removed by catastrophic 
losses, EPs are subject to energy losses which modify the shape of 
their distribution function.  Whether coulombian, synchrotron, inverse 
Compton, adiabatic or of any kind, these energy losses can generally 
be expressed by means of an \emph{energy loss function}, $\d E/\d t = 
\dot{E}_{i}(E,\vec{r},t)$, giving the energy loss rate ($\dot{E}_{i} < 
0$) of nuclei $i$ (in $(\mathrm{MeV/n})\mathrm{s}^{-1}$) as a function 
of energy, location and time.  Note that the energy loss function can 
also include positive terms corresponding to (non diffusive) steady 
acceleration.

The last physical ingredient that we need to know is the diffusion 
coefficient $D_{i}(E,\vec{r},t)$.  Together with 
$q_{i}(E,\vec{r},t)$, $\tau_{i}^{\mathrm{tot}}(E,\vec{r},t)$ and 
$\dot{E}_{i}(E,\vec{r},t)$, it allows one to calculate the 
distribution function of each EP species, $n_{i}(E,\vec{r},t)$, at 
least in principle.  The rate of any given reaction at any place and 
time is then obtained by integrating the relevant cross section over 
$n_{i}(E,\vec{r},t)$ in the energy space, just as in 
Eq.~(\ref{SecondaryProductionRate}), where the index $k$ may represent 
any secondary nucleus or photon species (e.g.  from a given gamma-ray 
line) in which we are interested.

The above ingredients being assumed known, we can now write down the 
propagation equation and its solution in a few general cases.

\section{One-zone models with stationary propagation conditions}
\label{OneZoneOnePhase}

In a number of cases, it is useless to study the spatial diffusion of 
the EPs, not because it is negligible, but because we are interested 
in the gamma-ray emission or secondary nuclei production integrated 
over a sufficiently large volume to be thought of as a box without 
internal structure.  A `one-zone model' can then be used, in which the 
space coordinates are purely and simply dropped, and the EP 
distribution function, $n_{i}(E,\vec{r},t)$, is replaced by its 
integral over the volume, namely the spectral density, $N_{i}(E,t)$, 
defined as the differential number of EPs at energy $E$ and time $t$ 
(in $(\mathrm{MeV/n})^{-1}$) in the whole box.  For example, such 
models are appropriate for the evaluation of reaction rates in a 
homogeneous medium, or more generally in an instrumentally unresolved 
region of space.

In a one-zone model, we have to solve the EP propagation equation in 
the following form~:
\begin{equation}
\begin{split}
	\frac{\partial}{\partial t}N_{i}(E,t) &+ \frac{\partial}
	{\partial E}(\dot E_{i}(E) N_{i}(E,t) )\\
	&= Q_{i}(E,t) + Q_{i}^{\prime}(E,t) -
	\frac{N_{i}(E,t)}{\tau_{i}^{\mathrm{tot}}},
	\label{PropEqOneZoneOnePhase}
\end{split}
\end{equation}
which merely expresses that the rate of change of the spectral density 
is equal to what is injected minus what is lost.  $Q_{i}(E,t)$ and 
$Q_{i}^{\prime}(E,t)$ are the integrals of $q_{i}(E,\vec{r},t)$ and 
$q_{i}^{\prime}(E,\vec{r},t)$ (i.e.  
Eq.~(\ref{SecondaryProductionRate})) over the volume of the box, and 
the second term in the left hand side describes the drift in the 
energy space, due to energy losses.

Let us now distinguish between primary EPs, for which $Q_{i}(E,t) \gg 
Q_{i}^{\prime}(E,t)$, and secondary EPs, for which $Q_{i}(E,t)$ $\ll 
Q_{i}^{\prime}(E,t)$.  Primary EPs are typically protons, alpha 
particles and abundant metals such as C, N, O, Mg, Si, Fe, etc., whose 
production rates by spallation are usually much smaller than their 
injection rate from the acceleration of ambient materials.  Energetic 
electrons are also mainly primary EPs.  On the contrary, nuclei such 
as Li, Be and B, as most of the usual spallation products, are 
secondary EPs.  This is also the case of positrons and anti-protons, 
both of which may be important tracers of cosmic-ray propagation.  
Such a distinction simplifies the calculation both conceptually and 
mathematically.  Indeed, we shall first solve the propagation equation 
for the primary nuclei, with the term $Q_{i}^{\prime}(E,t)$ dropped, 
and then use the spectral densities obtained in this ways to calculate 
the production rates of secondary nuclei.  For EP species with 
$Q_{i}(E,t)$ and $Q_{i}^{\prime}(E,t)$ of the same order of magnitude, 
one has to solve first the propagation equation for the parent nuclei, 
deduce $Q_{i}^{\prime}(E,t)$ and then solve 
Eq.~(\ref{PropEqOneZoneOnePhase}) again, in a (hopefully quickly 
convergent) iterative way.

The formal solution of Eq.~(\ref{PropEqOneZoneOnePhase}) reads~:
\begin{multline}
	N_{i}(E,t) = \frac{1}{|\dot{E}_{i}(E)|}\int_{E}^{+\infty}
	Q_{i}(E_{0},t - \tau_{i}(E_{0},E))\\
	\times \exp\Big(-
	\int_{E_{0}}^{E}\frac{\d\Ep}{\dot{E}_{i}(\Ep)
	\tau_{\mathrm{tot},i}(\Ep)}\Big)\d E_{0},
	\label{FormalSolution}
\end{multline}
where the function $\tau_{i}(E_{0},E)$ has an important physical 
meaning.  It is defined for each EP species $i$ as~:
\begin{equation}
	\tau_{i}(E_{0},E) = \int_{E_{0}}^{E}\frac{\d
	\Ep}{\dot{E_{i}}(\Ep)},
	\label{TauFonction}
\end{equation}
so it has the dimensions of a time, and describes the drift in energy 
of the EPs, due to the various energy loss mechanisms, operating at a 
rate $\dot{E_{i}}$.  We shall call $\tau$ the energy \emph{drift 
function}, or \emph{drift time}.  It is clear from 
Eq.~(\ref{TauFonction}) that $\tau_{i}(E_{0},E)$ is nothing but the 
time needed by a particle of species $i$ to slow down from energy 
$E_{0}$ to energy $E$.  This makes the formal solution, 
Eq.~(\ref{FormalSolution}), very easy to understand.  For a particle 
having a given energy, $E$, at time $t$, the drift function sets a 
one-to-one relation between the time at which it has been injected in 
the box, and the energy it then had.  Eq.~(\ref{FormalSolution}) thus 
merely says that the particles found at time $t$ at energy $E$ are the 
collection (the mathematical sum) of all the particles injected at a 
higher energy, $E_{0}\ge E$, but at an earlier time, $t_{0}$, such 
that since this time, their energy has dropped exactly to energy $E$, 
as the result of the various energy loss mechanisms.  The argument 
$t_{0} = t - \tau_{i}(E_{0},E)$ can thus be interpreted as a mere 
\emph{retarded time}, expressing the delay between the injection at 
$(E_{0},t_{0})$ and the collection at $(E,t)$ .

As for the exponential factor in Eq.~(\ref{FormalSolution}), it is 
equal to one in the absence of catastrophic losses 
($\tau_{\mathrm{tot}} \rightarrow \infty$), and otherwise corrects the 
contribution of the EPs injected at energies $E_{0}\ge E$ by weighting 
it by the probability of survival during the time needed to slow down 
from energy $E_{0}$ to energy $E$, that is during the drift time 
$\tau_{i}(E_{0},E)$.  This survival probability is $\mathcal{P}_{i} = 
\exp(-\langle \tau_{i}/\tau_{\mathrm{tot}}\rangle)$, where the average 
accounts for the energy dependence of catastrophic losses, 
$\tau_{\mathrm{tot}}(\Ep)$, which are indeed varying during the energy 
drift.  The time spent by the particles between energies \Ep~and $\Ep 
+ \d\Ep$ being $\d t(\Ep) = \d\Ep/\dot{E}_{i}(\Ep)$, one gets~:
\begin{equation}
	\left\langle \frac{\tau_{i}}{\tau_{\mathrm{tot}}}\right\rangle = 
	\int_{E_{0}}^{E}\frac{\d\Ep}{\dot{E}_{i}(\Ep) 
	\tau_{\mathrm{tot},i}(\Ep)},
	\label{MeanTauOverTauTot}
\end{equation}
which justifies the form of the exponential factor in 
Eq.~(\ref{FormalSolution}).

\section{One-zone models with time-dependent conditions of propagation}
\label{OneZoneSeveralPhases}

In the previous section we assumed that the energy loss rate, 
$\dot{E}_{i}(E)$, depended on the EP energy, but not on time.  Now let 
us consider the case of ionization losses in a medium of density 
$\rho_{0}$, whose chemical composition is given by the relative 
abundance (by number) of each nuclear species, $j$: $\alpha_{j}$.  Let 
$\d E/\d x|_{j}$ be the particle energy loss per unit grammage passed 
through in a medium of pure $j$ nuclei (expressed in g/cm$^{2}$).  
Then the energy loss rate of particles $i$ in the ISM is given by~:
\begin{equation}
	\dot{E}_{i}(E) = \frac{\d E}{\d x}\bigg|_{j}\times\alpha_{j}\rho_{0}v.
\end{equation}
As a consequence, the EP energy loss rate cannot be considered 
constant whenever the density or composition of the propagation medium 
are functions of time.  Now such situations are not unusual in 
astrophysics.  One only has to think of the medium surrounding an 
active massive star: it is composed of successive layers of different 
density and composition, resulting from successive phases of stellar 
wind (main sequence, red supergiant, Wolf-Rayet of type N, C, etc.).  
Such a situation has been considered by Parizot et~al.  (1997a,b,c).  
Likewise, EPs interacting in the interior of a supernova remnant will 
experience variations in the ambient density and composition (due to 
the expansion of the remnant and the dilution of the ejecta by the 
swept-up material).  This situation is addressed in detail in Parizot 
\& Drury (1999a,b).

\begin{figure*}
\resizebox{\hsize}{!}
{\includegraphics{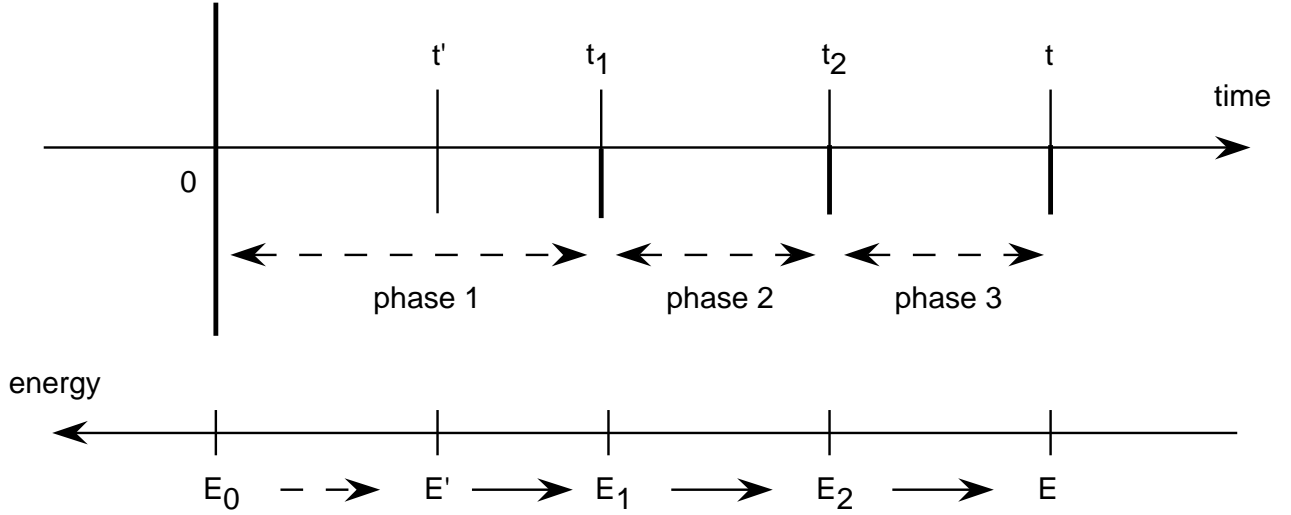}}
\caption{Diagram corresponding to an EP propagation process divided 
into three phases, during which the ambient density and chemical 
composition, and thus the energy loss rate, are assumed constant.  
Instants $t_{1}$ and $t_{2}$ correspond to the changes of phase, and 
$t$ is the `current' time, at which we want to calculate the EP energy 
spectrum.  The energy axis goes backward in time, because the energy 
losses produce a continual decrease of the energy of particles.  
Particles having energy $E$ at time $t$ (whose density, $N(E,t)$, we 
are looking for), had energy $E_{2}$ at time $t_{2}$, $E_{1}$ at time 
$t_{1}$, and $E_{0}$ at time $t_{0} = 0$.  These limiting energies of 
course depend on $E$ and $t$.  The diagram can be easily generalized 
to as much phases as one wishes.  The contribution of all phases to 
the current spectral density must be added, that of phase $i$ being 
limited to a domain of energy stretching from $E_{i-1}$ to $E_{i}$ 
(see text).}
\label{phases}
\end{figure*}

To deal with this non-stationarity of the conditions of propagation, 
we divide the process into successive phases during which the density 
and composition of the target can legitimately be considered constant.  
The solution of the propagation equation restricted to any individual 
phase is obtained in exactly the same way as above 
(Eq.~(\ref{FormalSolution})), and the only technical difficulty lies 
in the proper connection between the successive stationary solutions.  
To see how this works, let us consider the situation of 
Fig.~\ref{phases}.

We are interested in the EP distribution function at a succession of 
times $t$, during three distinct phases following the onset of the EP 
injection process, assumed to occur at $t = 0$ (by definition, the 
ambient density and composition are constant during each of these 
phases).  The set of these distribution functions would typically 
allow one to calculate observables such as the gamma-ray line emission 
rate as a function of time.  Let us first calculate the distribution 
function during phase~1, i.e for $0\le t\le t_{1}$, where $t_{1}$ 
marks the end of the first phase.  Since the conditions of propagation 
are constant during each phase, the EP distribution function at any 
time before $t_{1}$ is readily given by Eq.~(\ref{FormalSolution}).  
Dropping the species index, $i$, for convenience and noting 
$\dot{E}_{1}(E)$ and $\tau_{1}(E_{0},E)$ the energy loss function and 
the energy drift function during phase~1, we get~:
\begin{equation}
\begin{split}
	N(E,t) &= \frac{1}{|\dot{E}_{1}(E)|}\\
	&\times\int_{E}^{\infty}\d \Ep
	Q(\Ep,t - \tau_{1}(\Ep,E))\mathcal{P}_{1}(\Ep,E),
	\label{N1}
\end{split}
\end{equation}
where $\mathcal{P}_{1}(\Ep,E) = \exp(-\langle 
\tau_{1}/\tau_{\mathrm{tot},1}\rangle)$ is the probability (calculated 
under the conditions of phase~1) for an EP injected at energy \Ep~to 
reach energy $E$ without escaping or being altered by nuclear 
reactions (see Eq.~(\ref{MeanTauOverTauTot})).

Note that since the injection of EPs began at time $t = 0$, the 
injection rate is null at negative times, so that $Q(\Ep,t - 
\tau_{1}(\Ep,E))$, considered as a function of the injection energy, 
\Ep, vanishes for all energies greater than a maximum value, $E_{0}$, 
defined by~:
\begin{equation}
	\tau_{1}(E_{0},E) = t.
	\label{E0Definition}
\end{equation}
Indeed, any particle injected at a higher energy would need a time 
greater than $t$ to slow down to energy $E$, and consequently cannot 
contribute to $N(E,t)$.  The upper limit in the integral of 
Eq.~(\ref{N1}) can thus be replaced by the energy $E_{0}(E,t)$, 
implicitly defined in Eq.~(\ref{E0Definition}).  In order to shorten 
the writing of equations, we define the \emph{retarded injection 
function}, $f_{1}(\Ep,E)$, as~:
\begin{equation}
	f_{1}(\Ep,E) = Q(\Ep,t - \tau_{1}(\Ep,E))\times\mathcal{P}_{1}(\Ep,E),
	\label{RetardedInjectionFunction1}
\end{equation}
whose physical meaning is straightforward, so that Eq.~(\ref{N1}) can 
be re-written as~:
\begin{equation}
	N(E,t) = \frac{1}{|\dot{E}_{1}(E)|}\int_{E}^{E_{0}(E,t)}
	\hspace{-6pt}f_{1}(\Ep,E)\,\d\Ep.
	\label{N1Abbrev}
\end{equation}

Let us now calculate the EP spectral density, $N(E,t)$, during 
phase~2, i.e.  at times $t_{1}\le t\le t_{2}$.  One can still define a 
limiting energy, $E_{1}(E,t)$, such that the particles having this 
energy at time $t_{1}$ (end of phase~1, beginning of phase~2) have 
exactly energy $E$ at time $t$.  $E_{1}(E,t)$ is defined similarly to 
$E_{0}$ in Eq.~(\ref{E0Definition}), except that the energy drift 
function, $\tau$, now corresponds to the specific propagation 
conditions prevailing during phase~2~:
\begin{equation}
	\tau_{2}(E_{1},E) = t - t_{1}.
	\label{E1Definition}
\end{equation}
As above, any particle injected at an energy $\Ep\ge E_{1}$ 
cannot contribute to the spectral density $N(E,t)$ unless it has been 
injected earlier than $t_{1}$, i.e.  during phase~1.

Had the injection of EPs began at time $t_{1}$, we could easily write 
the solution of the propagation equation during phase~2, just as 
above~:
\begin{equation}
	N(E,t) = \frac{1}{|\dot{E}_{2}(E)|}\int_{E}^{E_{1}(E,t)}
	f_{2}(\Ep,E)\d\Ep,
	\label{N2}
\end{equation}
where the retarded injection function is now
\begin{equation}
	f_{2}(\Ep,E) = Q(\Ep,t - \tau_{2}(\Ep,E))
	\times\mathcal{P}_{2}(\Ep,E),
	\label{RetardedInjectionFunction2}
\end{equation}
and $\tau_{2}(\Ep,E)$ and $\mathcal{P}_{2}(\Ep,E)$ are the new energy 
drift function and EP survival probability, acknowledging the new 
target density and composition (i.e.  corresponding to phase~2).  But 
to obtain the actual EP spectral density, we still have to add the 
contribution of all the EPs injected during phase~1.  Now by 
definition of $E_{1}(E,t)$, $N(E,t)$ collects at time $t$ all the EPs 
which had energy $E_{1}(E,t)$ at the end of phase~1, i.e.  at time 
$t_{1}$.  But their number per unit energy, $N(E_{1}(E,t),t_{1})$, has 
already been calculated.  It is given by Eq.~(\ref{N1}) (or its 
condensed version Eq.~(\ref{N1Abbrev}))~:
\begin{equation}
\begin{split}
	N(E_{1}(E,t),t_{1}) &= \frac{1}{|\dot{E}_{1}(E_{1}(E,t))|}\\
	&\times\int_{E_{1}(E,t)}^{E_{0}(E,t)}f_{1}(\Ep,E_{1}(E,t))\d\Ep,
	\label{N1(t1)}
\end{split}
\end{equation}
where we recognized that
\begin{equation}
	E_{0}(E_{1}(E,t),t_{1}) = E_{0}(E,t).
	\label{E0EqualsE0}
\end{equation}

Before adding this contribution from phase~1 to the spectral density 
of the EPs at time $t$, we still need to take into account the 
catastrophic losses of particles since the end of phase~1.  This is 
achieved by multiplying $N(E_{1}(E,t),t_{1})$ by the survival 
probability (under the conditions of phase~2) from energy $E_{1}(E,t)$ 
to energy $E$.  Finally, we must remember that the distribution 
function $N(E,t)$ gives the differential number of EPs at energy $E$, 
that is the number of EPs between $E$ and $E + \d E$.  Accordingly, we 
must re-scale $N(E_{1}(E),t_{1})$ by a factor 
$|\dot{E}_{2}(E_{1}(E,t))|/|\dot{E}_{2}(E)|$ representing the 
contraction (or dilation) of the `comoving' energy interval.

Summing up all this `book keeping', we finally write the spectral 
density of the EPs at any time $t$ during phase~2 as~:
\begin{equation}
\begin{split}
	N(E,t) &= \frac{1}{|\dot{E}_{2}(E)|}\int_{E}^{E_{1}(E,t)}
	f_{2}(\Ep,E)\d\Ep\\
	&+ N(E_{1}(E,t),t_{1})\mathcal{P}_{2}(E_{1}(E,t),E)
	\frac{|\dot{E}_{2}(E_{1}(E,t))|}{|\dot{E}_{2}(E)|},
	\label{N2bis}
\end{split}
\end{equation}
or, substituting from Eq.~(\ref{N1(t1)}) and conveniently abbreviating 
$E_{i}(E,t)$ to $E_{i}$,
\begin{equation}
\begin{split}
	N(E,t) &= \frac{1}{|\dot{E}_{2}(E)|}\bigg[\int_{E}^{E_{1}}
	f_{2}(\Ep,E)\d\Ep\\
	&+ \frac{|\dot{E}_{2}(E_{1})|}{|\dot{E}_{1}(E_{1})|}
	\mathcal{P}_{2}(E_{1},E) \int_{E_{1}}^{E_{0}}
	f_{1}(\Ep,E_{1})\d\Ep \bigg].
	\label{N2ter}
\end{split}
\end{equation}

Note that we could have alternately obtained Eq.~(\ref{N2bis}) by 
integrating the propagation equation, (\ref{PropEqOneZoneOnePhase}), 
beginning at time $t_{1}$ and replacing the injection function 
$Q(E,t)$ by an effective injection function, $Q_{\mathrm{eff}}$, 
defined by~:
\begin{equation}
	Q_{\mathrm{eff}}(E,t) = Q(E,t) + N(E,t_{1})\delta(t - t_{1}).
	\label{Qeff}
\end{equation}
Although mathematically equivalent, we feel however that the 
derivation given above gives a deeper physical insight of the 
situation.

We finish the study of the multiphase case by computing the EP 
spectral density during phase~3.  The calculation leading to 
Eq.~(\ref{N2ter}) generalizes straightforwardly if we introduce the 
energy, $E_{2}(E,t)$, that a particle of current energy $E$ (i.e.  at 
time $t$) had at the time $t_{2}$ marking the end of phase~2 and the 
beginning of phase~3.  As above, $E_{2}(E,t)$ is defined by~:
\begin{equation}
	\tau_{3}(E_{2},E) = t - t_{2},
	\label{E2Definition}
\end{equation}
where the subscript `3' refers to the conditions of propagation during 
phase~3.

Recursively adding the contributions of each phase to the spectral 
density at time $t$, we obtain~:
\begin{multline}
	|\dot{E}_{3}(E)|N(E,t) = \int_{E}^{E_{2}}f_{3}(\Ep,E)\d\Ep\\
	+ \frac{|\dot{E}_{3}(E_{2})|}{|\dot{E}_{2}(E_{2})|}
	\mathcal{P}_{3}(E_{2},E) \int_{E_{2}}^{E_{1}}f_{2}(\Ep,E_{2})\d\Ep\\
	+ \frac{|\dot{E}_{2}(E_{1})|}{|\dot{E}_{1}(E_{1})|}
	  \frac{|\dot{E}_{3}(E_{2})|}{|\dot{E}_{2}(E_{2})|}
	  \mathcal{P}_{2}(E_{1},E_{2})\mathcal{P}_{3}(E_{2},E)\\
	\times 
	  \int_{E_{1}}^{E_{0}}f_{1}(\Ep,E_{1})\d\Ep,
	\label{N3}
\end{multline}
with straightforward generalization to any number of phases.

\section{Extended models with homogeneous spatial diffusion}
\label{ExtendedModels}

Although one-zone models are both powerful and easy to handle, 
gamma-ray observations with high angular resolution might make it also 
necessary to use extended models, allowing one to study inhomogeneous 
astrophysical sites.  As an example, one may be interested in 
situations where energetic particles are accelerated in a small region 
of space and diffuse in the surrounding medium, exploring sites with 
various densities and magnetic properties.  The EP distribution 
function, and hence the gamma-ray emission, should thus be different 
in neighboring regions which can be resolved by the instruments.  
Moreover, as was discussed in Parizot (1997), the detailed study of 
spatial diffusion is sometimes indispensable, even in homogeneous 
media, if the loss of EPs by escape out of the confinement region has 
to be taken into account.

Let us then write the propagation equation of the EPs in the case of 
an extended model.  We are now interested in the EP distribution 
function, $n_{i}(E,\vec{r},t)$, rather than in the integrated spectral 
density, $N_{i}(E,t)$.  It satisfies an equation similar to 
Eq.~(\ref{PropEqOneZoneOnePhase}), with new terms describing the 
spatial transport of particles.  In a number of applications, 
convection can be neglected, because one can work in the referential 
of the interstellar plasma, or average over large enough regions of 
space for the plasma to be considered globally at rest.  Keeping only 
the diffusion term, we then have the EP propagation equation in the 
extended case~:
\begin{multline}
	\frac{\partial}{\partial t}n_{i}(\vec{r},E,t) + 
	\frac{\partial}{\partial E}(\dot{E}_{i}(E)n_{i}(\vec{r},E,t)) = 
	q_{i}(\vec{r},E,t)\\
	- \frac{n_{i}(\vec{r},E,t)}{\tau_{\mathrm{tot}}}
	+ \nabla(D(\vec{r},E,t)\nabla n_{i}(\vec{r},E,t)).
	\label{PropagEquaExtendedModel}
\end{multline}

This equation can of course be solved numerically, but it is much 
better (be it only regarding computation time) to solve it formally 
once for all, for any injection function $q_{i}(\vec{r},E,t)$, as in 
the one-zone case.  To do this, however, we shall assume that the 
diffusion coefficient does actually not depend on the location, 
$\vec{r}$, and time, $t$.  On the other hand, it can still depend on 
(and be any function of) the energy, as has to be the case in most 
realistic astrophysical situations.  Note that although the condition 
$D(E,\vec{r},t) = D(E)$ may be too restrictive in some specific 
occasions, there will be a way around it in most cases.  Indeed the 
diffusion conditions (gathered mathematically in the diffusion 
coefficient, $D$) do not generally change smoothly, but rather 
sharply, while passing from one type of environment to another, like 
from the ISM to a dense cloud or from a superbubble to the ISM. In 
such cases, one can use a procedure analogous to that of 
Sect.~\ref{OneZoneSeveralPhases} and solve the propagation equation 
locally, in regions with homogeneous diffusion, and connect the 
solutions at the boundaries in much the same way as we did above in 
the multi-phase case.  In practical terms, this amounts to adding to 
the injection function, $q_{i}(E,\vec{r},t)$, a term of the form 
$\int_{S}\mathrm{\delta}(\vec{r} - \vec{r_{0}})\d\vec{S}$, 
corresponding to the flux of particles across the boundary of each 
region.  This flux `inherited' from the propagation of EPs in 
neighboring regions is exactly analogous to the spectral density 
`inherited' from the propagation during the earlier phase, as we 
encountered it in Sect.~\ref{OneZoneSeveralPhases}.

The solution of Eq.~(\ref{PropagEquaExtendedModel}) (with homogeneous 
diffusion) can be given in a convenient way by introducing a new 
function, $\chi_{i}(E_{0},E)$, in the spirit of the energy drift 
function given by Eq.~(\ref{TauFonction})~:
\begin{equation}
	\chi_{i}(E_{0},E) = \int_{E_{0}}^{E}D(\Ep)
	\frac{\d\Ep}{\dot{E}(\Ep)}.
	\label{FunctionChi}
\end{equation}
Physically, $\chi_{i}(E_{0},E)$ represents the effective value of 
`$Dt$' when the particle slows down from energy $E_{0}$ (at which it 
has been injected) to energy $E$ (which it has at time $t$).  This is 
easily seen by noting that $\d\Ep/\dot{E}(\Ep)$ represents the 
time passed by the particle between energies $\Ep + \d\Ep$ and $\Ep$.  
The introduction of this `effective' diffusion parameter is necessary 
because the energy of the particle decreases with time, while the 
diffusion coefficient depends on energy.  In the case when $D(E) = 
D_{0}$ is constant, we have in fact $\chi_{i}(E_{0},E) = 
D_{0}\tau_{i}(E_{0},E)$.  Note finally that $\chi_{i}(E_{0},E)$ has 
the dimensions of a surface, so we shall call it the \emph{diffusion 
section} of the particle (when passing from $E_{0}$ to $E$).

We can now give the EP distribution function at energy $E$, location 
$\vec{r}$ and time $t$, in the case of a homogeneous, though energy 
dependent diffusion coefficient~:
\begin{multline}
	n_{i}(E,\vec{r},t) =
	\frac{1}{|\dot{E}_{i}(E)|}\times\\
	\int_{E}^{\infty}\d\Ep
	\int\d\vec{r}^{\prime}\,f_{i}(\vec{r}^{\prime};\Ep,E)
	g_{i}(\vec{r}^{\prime},\vec{r};\Ep,E),
	\label{ExtendedModelFormalSolution}
\end{multline}
where $f_{i}$ is the 3D generalization of the retarded injection 
function (Eq.~(\ref{RetardedInjectionFunction1}))~:
\begin{multline}
	f_{i}(\vec{r}^{\prime};\Ep,E) =
	q_{i}(\Ep,\vec{r}^{\prime},t - \tau_{i}(\Ep,E))\\
	\times\exp\Big(-
	\int_{\Ep}^{E}\frac{\d\Epp}{\dot{E}_{i}(\Epp)
	\tau_{\mathrm{tot},i}(\Epp)}\Big),
	\label{3DRetardedInjectionFunction1}
\end{multline}
and $g_{i}$ is given by~:
\begin{equation}
	g_{i}(\vec{r}^{\prime},\vec{r};\Ep,E) = 
	\frac{1}{8(\pi\chi(\Ep,E))^{3/2}}
	\exp\left(
	-\frac{\|\vec{r} - \vec{r}^{\prime}\|^{2}}{4\chi(\Ep,E)}\right).
	\label{DiffusionSurvivalProba}
\end{equation}

This last function proves to have a very simple physical 
interpretation.  To see this, let us consider the simplest equation of 
diffusion, for identical particles in a homogeneous and isotropic 
medium, $\partial N/\partial t - D \Delta N = 0$, where $N$ is the 
number density of particles, $D$ is the diffusion coefficient, and 
$\Delta$ denotes the Laplacian operator.  In an infinite medium, if 
$N_{0}$ particles are initially injected at $\vec{r} = 
\vec{r}_{\mathrm{in}}$, then we have the well-known solution~:
\begin{equation}
	N(\vec{r},t) = \frac{N_{0}}{8(\pi Dt)^{3/2}}
	\exp\left(- \frac{\|\vec{r} - \vec{r}_{\mathrm{in}}\|^{2}}{4Dt}\right).
\end{equation}
One way to look at this solution, emphasizing the stochasticity of the 
diffusion process, is to say that if a particle is introduced in 
$\vec{r}_{\mathrm{in}}$ at $t = 0$, the density of probability that it 
be in $\vec{r}$ at time $t$ is 
$\mathcal{P}_{t}(\vec{r}_{\mathrm{in}},\vec{r}) = N(\vec{r},t)/N_{0}$.  
Now comparing the expressions for 
$\mathcal{P}_{t}(\vec{r}_{\mathrm{in}},\vec{r})$ and 
$g(\vec{r}^{\prime},\vec{r};\Ep,E)$ (given by 
Eq.~(\ref{DiffusionSurvivalProba})), while remembering that 
$\chi(\Ep,E)$ is the effective `Dt' during the time it takes the 
particles to slow down from $\Ep$ to $E$, it is clear that
\begin{equation}
	g_{i}(\vec{r}^{\prime},\vec{r};\Ep,E) =
	\mathcal{P}_{\tau(E^{\prime},E)}(\vec{r}^{\prime},\vec{r}),
	\label{g=P}
\end{equation}
or, to say it in words, the function 
$g_{i}(\vec{r}^{\prime},\vec{r};\Ep,E)$ appearing in 
Eq.~(\ref{ExtendedModelFormalSolution}) is nothing but the probability 
for a particle injected at location $\vec{r}^{\prime}$ at energy $\Ep$ 
to be found in $\vec{r}$ at energy $E$.

With the definition $h_{i} = f_{i}\times g_{i}$, we can now re-write 
the solution (\ref{ExtendedModelFormalSolution}) of the propagation 
equation in the simplest way~:
\begin{equation}
	n_{i}(E,\vec{r},t) = \frac{1}{|\dot{E}_{i}(E)|}
	\int_{E}^{\infty}\d\Ep\int\d\vec{r}^{\prime}\,
	h_{i}(\vec{r}^{\prime},\vec{r};\Ep,E),
	\label{ExtendedModelFormalSolution2}
\end{equation}
where $h_{i}$ should be thought of as the `full' retarded injection 
function, i.e.  retarded in both space and time (factors $g_{i}$ and 
$f_{i}$, respectively).  The physical interpretation of the solution 
is then obvious: the particles which one finds at energy $E$, time $t$ 
and location $\vec{r}$, are the collection of all the particles which 
have been injected at a higher energy $\Ep$, anywhere in the volume 
considered (integration over $\vec{r}^{\prime}$), but at a time in the 
past such that, since their injection, the energy losses have brought 
their energy down from $\Ep$ to exactly $E$.  In addition, each of 
these contributions must be weighted by the probability that the 
particle survived during this time (exponential factor in 
Eq.~(\ref{3DRetardedInjectionFunction1})) and that, during this time 
again, it diffused from its injection site, $\vec{r}^{\prime}$, to the 
place now probed, in $\vec{r}$ (function $g_{i}$, 
Eq.~(\ref{DiffusionSurvivalProba})).

\section{The stationary limit of the models}
\label{SteadyState}

As was discussed in the introduction, some astrophysical situations 
need not be studied in a time-dependent scheme.  In steady-state, 
after a transitory regime lasting about the time-scale of energy 
losses, the spectral density of the EP species $i$ reaches its 
equilibrium value, $N_{i}(E)$, which satisfies~:
\begin{equation}
	\frac{\partial}{\partial E} (\dot{E}_{i}(E)N_{i}(E)) = 
	Q_{i}(E) - \frac{N_{i}(E)}{\tau_{\mathrm{tot}}}.
	\label{StatPropEqua}
\end{equation}
Integrating this equation directly or, equivalently, taking the 
stationary limit of the general, time-dependent solution given above 
(Eq.~(\ref{FormalSolution})), one obtains~:
\begin{equation}
	N_{i}(E) = \frac{1}{|\dot{E}_{i}(E)|}\int_{E}^{+ \infty}
	Q_{i}(E_{0})\mathcal{P}_{i}(E_{0},E)\d E_{0},
	\label{StatSolution}
\end{equation}
where the survival probability $\mathcal{P}_{i}$ has the same 
expression as above.

It is important to note that, in the steady-state approximation, the 
spectral density of stable energetic particles is inversely 
proportional to the ambient density, provided that the escape can be 
neglected, i.e.  for low energy particles or in the case of a thick 
target.  Indeed, the Coulombian energy loss rate, $\dot{E}_{i}(E)$, is 
always proportional to the ambient density, $n_{0}$, while the 
catastrophic loss time, $\tau_{\mathrm{tot}}$, is then equal to the 
nuclear destruction time, $\tau_{\mathrm{D}}$, which according to 
Eq.~(\ref{destructionTime}), behaves as $n_{0}^{-1}$.  As a result, 
the survival probability, $\mathcal{P}_{i}$, is independent of 
$n_{0}$, and inspection of Eq.~(\ref{StatSolution}) shows that 
$N_{i}(E)$ scales as $n_{0}^{-1}$.  Now the nuclear reaction rates, 
due to the interaction of the EPs with the ISM, are obtained by 
integrating the cross-sections over the spectral density~:
\begin{equation}
	Q_{k} = \sum_{i,j}\int_{0}^{+\infty}n_{j}N_{i}(E) 
	\sigma_{ij\rightarrow k}(E)v_{i}(E)\d E,
	\label{TauxProdStat}
\end{equation}
where the notations are the same as in 
Eq.~(\ref{SecondaryProductionRate}).

It is clear, then, that the $n_{0}^{-1}$ dependence of $N_{i}$ is 
exactly compensated by the factor $n_{j}$, which is just $n_{0}$ times 
the relative abundance of nuclei $j$ in the ISM. As a conclusion, 
\emph{all the nuclear reaction rates are independent of density in the 
steady-state approximation}.  This is worth emphasizing, because it is 
one fundamental phenomenological distinction between steady-state and 
time-dependent models.  In essence, steady-state models cannot 
acknowledge density dependences.  This is no longer true for 
time-dependent models, as we have shown elsewhere on various examples 
(Parizot et~al.  1997a,b,c; Parizot \& Drury 1999a,b).

\section{Conclusion}

In conclusion, we have reviewed the solutions of the differential 
equation describing the propagation of energetic particles (EPs) in 
the ISM, in the case of both one-zone and extended models, taking into 
account homogeneous spatial diffusion.  The solutions obtained admit a 
simple physical interpretation, which we emphasized introducing 
intermediate functions~: the energy drift function, 
$\tau_{i}(E_{0},E)$, and the diffusion section, $\chi_{i}(E_{0},E)$, 
which both depend on the nuclear species of the EPs.  The first one 
describes the time it takes a particle to slow down from its injection 
energy, $E_{0}$, to its current energy, $E$, because of energy 
dependent energy losses.  As for the latter, it plays the role of an 
effective `diffusion coefficient-times-duration', taking into account 
the energy dependence of the diffusion coefficient.  We have used 
these functions to construct \emph{retarded injection functions} which 
sum up the energy history of the EPs and allow one to write the 
solution of the propagation equation in a simple and intelligible way.

We also studied the case when the conditions of propagation vary, 
which arises when either the composition or the density of the ambient 
medium are not constant.  We have shown how it is possible to use the 
standard results in this case, dividing the process into as much 
successive phases as required so that the conditions of propagation 
can be considered constant during each phase.  We outlined the 
detailed procedure for fitting the phases altogether, relying on, and 
emphasizing again, the physical content of the mathematics involved.  
The procedure can be used to solve any type of time-dependent 
astrophysical situation involving the interaction of energetic 
particles in the interstellar medium.  Indeed, the formal solutions 
derived here can be easily integrated to obtain the distribution 
function or spectral density of the EPs, from which the nuclear or 
electromagnetic reaction rates are deduced directly from the knowledge 
of the cross sections involved.

Finally, we considered the stationary limit of the models, stressing 
that the assumptions inherent in steady-state make it impossible 
\emph{in principle} to investigate the influence of the ambient 
density on the processes under study.  This is one crucial superiority 
of the time-dependent models which, for this reason, cannot be avoided 
in a great number of astrophysical situations.  As an example, Parizot 
et~al.  (1997a,b,c) have studied the time-dependent gamma-ray line 
emission induced by the winds of massive stars, and Parizot \& Drury 
(1999a,b) calculated the spallative nucleosynthesis within an 
expanding supernova remnant, taking the dynamics of the process into 
account.  In each case, the influence of the ambient density has been 
investigated, and in each case it proved very significant.  We believe 
that the improvement of the spatial resolution of gamma-ray detectors 
as well as the accumulation of data on variable gamma-ray sources will 
make the use of time-dependent models more and more necessary.

In the same spirit, as the Galactic chemical evolution models are 
getting more and more precise, the question should be raised whether 
steady-state models are still appropriate in all the situations.  In 
particular, recent studies of the evolution of the spallation products 
Li, Be and B, have focused mainly on the very early stages of Galactic 
evolution, when the metallicity of the ISM was less than one 
thousandth of the solar metallicity.  Depending on the models, this 
corresponds to an age of the order of the confinement or the energy 
loss time scales.  As a consequence, one should be careful in assuming 
steady-state, because the transitory regime may have not yet totally 
decreased.  In addition, the injection of cosmic rays is usually 
assumed to follow closely the supernova activity in the Galaxy, but it 
should be kept in mind that the supernova explosion rate may have 
changed on a very small time scale in the early Galaxy, which would be 
one more reason to be careful with steady-state models, and to 
investigate possible significant time-dependent effects.

\begin{acknowledgements}
I wish to thank Roland Lehoucq warmly for stimulating and much 
neuron-enjoyable discussions, and Luke Drury for his valuable comments 
and suggestions.  This work was supported by the TMR programme of the 
European Union under contract FMRX-CT98-0168.
\end{acknowledgements}

\end{document}